%% file: main.tex
\documentclass{article}

\usepackage[preprint]{neurips_2026}
\workshoptitle{ML for Systems}

\usepackage[utf8]{inputenc}
\usepackage[T1]{fontenc}
\usepackage{hyperref}
\hypersetup{hidelinks}
\setcitestyle{numbers,square}
\usepackage{url}
\usepackage{booktabs}
\usepackage{amsfonts}
\usepackage{amsmath}
\usepackage{amssymb}
\usepackage{nicefrac}
\usepackage{microtype}
\usepackage{xcolor}
\usepackage{enumitem}
\usepackage{tikz}
\usetikzlibrary{arrows.meta,positioning}
\setlist{nosep,leftmargin=*}
\usepackage{graphicx}
\usepackage{float}

\title{Network-Aware Forecasting on Wireless Access Points}
\author{
  Niloo Bahadori \\
  Cisco Systems \\
  \texttt{nbahador@cisco.com}
  \And
  Swadhin Pradhan \\
  Cisco Systems \\
  \texttt{swapradh@cisco.com}
  \And
  Peiman Amini\\
  Cisco Systems \\
  \texttt{pamini@cisco.com}
}

\begin{document}

\maketitle

\begin{abstract}
Enterprise wireless access points (APs) are promising platforms for predictive machine learning (ML), but their primary responsibility remains providing wireless connectivity and network services. Predictive inference must therefore share an AP’s CPU and memory with packet processing, Wi-Fi and IoT radio operations, and client management. This resource contention creates two risks: a model that performs well on proxy hardware may be too slow on the target AP, while a model that fits in isolation may still degrade network services under load. We define \textit{network-aware deployability} using two gates: qualification of the model and its execution path on the target AP, followed by validation of its execution profile under packet-service and forecasting constraints. Our benchmarks show that edge testbeds do not reliably capture target behavior. Across matched artifacts and serving settings, five model implementations run 6.1--19.1$\times$ slower on an AP than on a Raspberry Pi~5, while peak memory usage differs by up to 22\%. Moreover, two forecasting foundation models of similar size differ in AP latency by 19$\times$. When serving a smaller model across 13 parallel streams at a 30~s cadence under network saturation, default execution increases p99 round-trip time (RTT) by 76\% and reduces throughput by 7.06\%. Understanding these trade-offs is essential for live deployment if we aim to use APs for both networking and ML workloads.  


\end{abstract}

\input{sections/01_introduction}
\input{sections/02_approach}
\input{sections/03_evaluation}

\input{sections/04_discussion_conclusion}

\clearpage

\input{references}
\clearpage
\appendix

\end{document}

%% file: sections/01_introduction.tex
\section{Introduction}
\label{sec:introduction}
Enterprise access points (APs) are continuously operating network appliances at the enterprise edge. They provide client access; process and forward or tunnel traffic; execute Wi-Fi, IoT-radio, and control-plane functions; and manage connected clients. They also have direct visibility into client-onboarding events, local system health, and wired-uplink telemetry, making them useful location for forecasting wireless, networking, and system behavior. Prior work showed that network-native time-series foundation models (TSFMs) can forecast wireless telemetry and that a compact model can achieve sub-second inference on AP-class proxy hardware~\cite{apex}. We ask the follow-up deployment question: \emph{when can recurring inference run on an AP without interfering with the network it serves?}

An AP differs from a general-purpose edge computer, even with similar CPU and memory capacity, because uninterrupted network service takes priority over secondary workloads. AP's resources must first support processing and forwarding traffic, managing the radio and control plane, and serving connected clients. 
The compute headroom available for predictive inference is therefore workload-dependent and can shrink as traffic, client count, and control-plane activity increase. On-AP inference is useful only when it meets the required prediction cadence without increasing client latency or reducing throughput.



Our experiments show two independent failure modes. First, parameter count and proxy hardware latency do not reliably identify the production operating point. Second, even a resident model that completes an individual inference in under one second can perturb throughput and tail latency under contention. These observations motivate two deployment checks: qualify the complete model and execution path on the target AP before deployment, and then verify that the recurring multi-stream forecasting workload remains within packet-service and prediction-timing budgets as network load changes (as shown in Fig.~\ref{fig:system_model}). We therefore formulate on-AP forecasting as an \emph{asymmetric, network-aware co-location problem}, rather than as a one-time model-placement decision or fair resource sharing among peer inference workloads.
We make three contributions:
\begin{enumerate}
    \item  A two-gate deploy methodology that qualifies the complete model and execution path on the target AP and then certifies load-specific execution profiles against a deployment contract containing packet-service and forecast-deadline requirements.
    \item We perform a matched AP/Raspberry Pi~5 characterization showing that inference latency transfers differently across models and cannot be estimated using a fixed proxy-to-device scaling factor.
    \item We evaluate a target-feasible workload serving 13 forecast streams under network contention and show that the runtime configuration determines whether it meets the network-performance budget.
\end{enumerate}

%% file: sections/02_approach.tex
\section{Two-Gate Deployability}
\label{sec:deployability}

\begin{figure}[t]
    \centering
    \includegraphics[width=0.95\columnwidth]{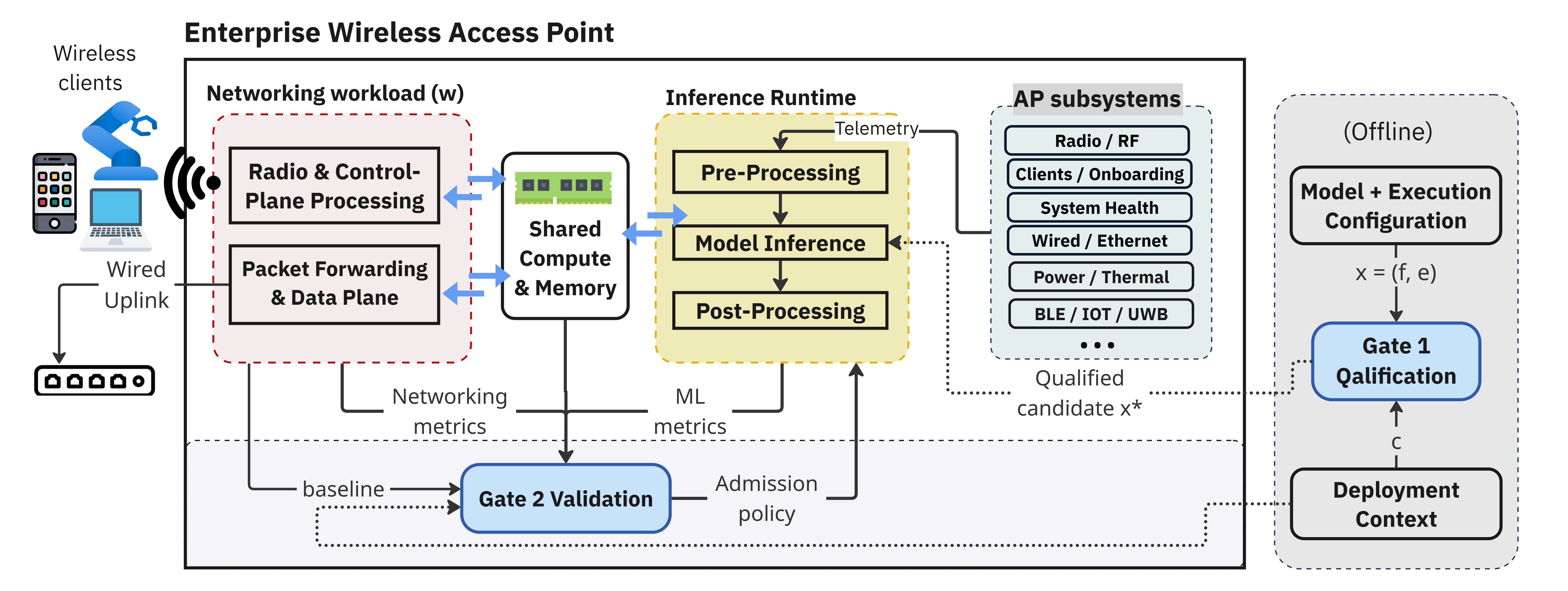}
    \caption{System model.}
    \label{fig:system_model}
\end{figure}

We express on-AP deployability as two feasibility gates over the complete model (shown in Fig.~\ref{fig:system_model}) and execution path, rather than as a single multi-objective optimizer. Let \(c\) denote the deployment context for a particular target environment. The context captures the target AP, application and
network-service class, and specifies the required forecast quality, resource budgets, acceptable forecast age, and tolerable packet-service degradation. For example, a latency-sensitive robotic-control WLAN may use a tighter packet-latency budget than a conventional office WLAN.
\paragraph{Gate 1: target-AP qualification.}
A candidate \(x=(f,e)\) combines a forecasting model \(f\) with an
execution configuration \(e\), including the artifact, precision,
runtime, threading, CPU placement and priority, and parallelism.
For deployment context \(c\), the candidate passes the target-AP gate
when
\begin{equation}
x \in \mathcal{X}_{\mathrm{AP}}(c)
\Longleftrightarrow
\begin{cases}
Q_i(x) \ge q_i^\star(c), & \forall i \in \mathcal{D},\\
M_{\mathrm{pk}}(x) \le B_M(c),\\
T_{\mathrm{cyc},0}^{99}(x,\mathcal{S}) \le D_{\mathrm{cyc}}(c).
\end{cases}
\label{eq:target_gate}
\end{equation}
Here, \(q_i^\star(c)\) is the required forecast quality for domain \(i\), \(B_M(c)\) is the memory budget available on the target AP, and \(D_{\mathrm{cyc}}(c)\) is the forecast-cycle deadline specified by the deployment context.

\paragraph{Gate 2: load-dependent operating envelope.}
For a networking workload \(w\) and a fixed execution profile
\(\pi\), let \(Y_0(w)\) and \(R_0(w)\) be throughput and p99 RTT
under a matched no-ML baseline, and let \(Y_\pi(w)\) and
\(R_\pi(w)\) be the corresponding measurements with recurring
inference. We report nonnegative degradation:
\begin{equation}
\begin{split}
d_{\mathrm{tp}}(w,\pi)
&=
\left[
\frac{Y_0(w)-Y_\pi(w)}{Y_0(w)}
\right]_+,\\
d_{\mathrm{rtt}}(w,\pi)
&=
\left[
\frac{R_\pi(w)-R_0(w)}{R_0(w)}
\right]_+,
\end{split}
\qquad [z]_+ \triangleq \max(z,0).
\label{eq:degradation}
\end{equation}
Thus, an observed throughput improvement or RTT reduction is reported
as zero degradation rather than as an ML-attributed service benefit.

A profile is admissible for workload \(w\) and deployment context
\(c\) when
\begin{equation}
\pi \in \Pi_{\mathrm{adm}}(w,c)
\Longleftrightarrow
\begin{cases}
d_{\mathrm{tp}}(w,\pi) \le \epsilon_{\mathrm{tp}}(c),\\
d_{\mathrm{rtt}}(w,\pi) \le \epsilon_{\mathrm{rtt}}(c),\\
T_{\mathrm{cyc}}^{99}(x,w,\pi,\mathcal{S})
    \le D_{\mathrm{cyc}}(c).
\end{cases}
\label{eq:runtime_gate}
\end{equation}
Forecast-accuracy thresholds and deadlines follow the required prediction cadence and validity interval of each use-case. Memory and compute budgets reflect the AP’s capacity after reserving resources for networking, while throughput and tail-latency limits follow network-service requirements.

%% file: sections/03_evaluation.tex
\section{Evaluation}
\label{sec:evaluation}

\paragraph{Setup.}
We evaluate a production-class, accelerator-free AP with a four-core ARM processor and approximately 2~GB of RAM. The resident candidate, APEX~2 (8.8M parameters), uses a 288 seconds long context sampled at 1~Hz and produces a point forecast plus nine quantiles. It runs every 30~s across 13 streams: client onboarding (2), AP self-health (5), and wired uplink (6). These domains collectively span three essential layers of the service path: access, local platform operation, and upstream connectivity. Each exhibits distinct dynamics, predictability, and error costs. Yet all compete for the same limited compute and memory resources.
For qualification, each model uses the same artifact and serving configuration on the AP and Raspberry Pi~5. Peak RSS and latency include the unoptimized proof-of-concept runtime and temporary tensors; therefore, the absolute values do not represent final product footprints.

We evaluate the two deployment gates separately. Gate~1 measures model latency and memory on the target AP. The forecast quality of these models on wireless telemetry is established in \cite{apex},  here we only study the systems gates. For Gate~2, one associated wireless client generates sustained traffic while recurring inference serves all 13 streams. Under default execution, we sweep approximately 100~Mbps, 250~Mbps, and saturation, conducting three 60~s trials per condition. A separate saturation run compares default and service-prioritized execution against matched no-ML baselines. The impact of the network is quantified relative to these baselines using throughput and p99 RTT. Forecast serviceability is assessed using the full 13-stream cycle-completion time under both execution profiles and the number of deadline misses. A cycle violates the freshness requirement if it exceeds the deployed cadence, $D_{\mathrm{cyc}}=30\,\mathrm{s}$, chosen as a practical timescale for the use cases considered. We set a 5\% network degradation as engineering tolerance. Together, the two gates distinguish target-device feasibility in isolation from safe coexistence under network contention.
\begin{figure*}[!t]
    \centering
    \includegraphics[width=0.495\textwidth]{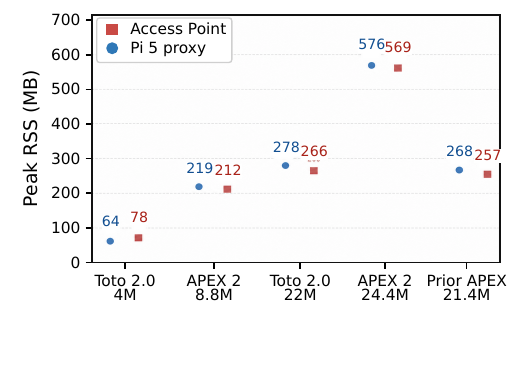}%
    \hfill
    \includegraphics[width=0.495\textwidth]{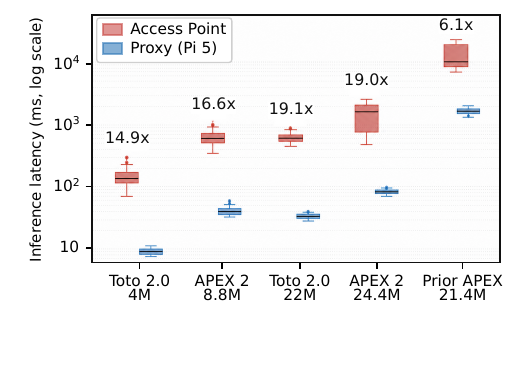}
    \vspace{-2mm}
    \caption{Target device characterization. Left: Peak RSS, Raspberry Pi~5 vs. access point. Right: Per forecasting task inference latency.}
    \label{fig:target_characterization}
    \vspace{-2mm}
\end{figure*}


\paragraph{Target hardware changes the feasible set.}
Figure~\ref{fig:target_characterization} shows strong model-dependent AP/Pi latency slowdowns (6.1--19.1$\times$), ruling out a fixed proxy-to-target scaling factor and showing that parameter count alone is insufficient. In contrast, peak RSS transfers closely (within 22\%), indicating that latency is dominated by model-platform interaction rather than coarse resource specifications.

\begin{table}[H]
\centering
\scriptsize
\setlength{\tabcolsep}{4pt}
\caption{Network Impact and Forecast Cycle of recurring APEX~2 inference on the production AP.} 
\label{tab:service}
\footnotesize

\textbf{(a) Network Impact under Increasing Load}

\vspace{5pt}
\begin{tabular}{lrr}
\toprule
Network load & Throughput loss & p99 RTT inflation \\
\midrule
$\sim$100 Mbps & 0\%    & 0\%  \\
$\sim$250 Mbps & 0\%    & 53\% \\
Saturation     & 7.06\% & 76\% \\
\bottomrule
\end{tabular}

\vspace{5pt}

\textbf{(b) Saturation: Service Protection versus Forecast Freshness}

\vspace{5pt}
\begin{tabular}{lrrrr}
\toprule
Profile &
Throughput loss &
p99 RTT inflation &
Cycle p99 &
Deadline misses \\
\midrule
Default 
    & 7.06\% & 76\% & 8.31\,s & 0/100 \\
Service-prioritized
    & 0\% & 0\%
    & 17.5\,s & 0/100 \\
\bottomrule
\end{tabular}
\end{table}

\paragraph{Network protection and inference speed.}
Under default ML execution, increasing network load first affects tail latency: at approximately 250~Mbps, p99 RTT increases by 53\% (no throughput loss). At saturation, the throughput loss reaches 7.06\% and the inflation of the p99 RTT reaches 76\%. 
Service-prioritized execution protects networking by restricting and deprioritizing inference, which leads to an admissible saturation operating point with no measured nonnegative network degradation. This protection slows the 13-stream forecast cycle from 8.31~s to 17.5~s p99, but both profiles remain below the deployed 30~s deadline with zero observed deadline misses. 
Our observations point to inference-runtime memory allocation and memory-system contention as key contributors to throughput loss and tail-latency inflation. However, because we do not isolate these mechanisms individually, we present this interpretation as a systems-level hypothesis rather than a component-level attribution.

%% file: sections/04_discussion_conclusion.tex
\section{Related Work}
\label{sec:related}
Platform-aware mobile systems establish that FLOPs and parameter counts poorly predict latency, motivating target-device measurement~\cite{mnasnet,netadapt}. HAQ uses hardware feedback for mixed-precision quantization~\cite{haq}, while ~\cite{onceforall} specializes subnetworks across devices and latency constraints. MCUNet co-designs models and inference software under tight edge constraints~\cite{mcunet}; DeepMon accelerates continuous vision on mobile GPUs~\cite{deepmon}; and NestDNN adapts competing DNNs to changing mobile resources~\cite{nestdnn}. Clockwork extends this systems perspective to predictable inference and request-level isolation in cloud model serving~\cite{clockwork}. These works enhance efficiency, match workloads to resources, or isolate inference. Enterprise APs instead prioritize forwarding, radio/control-plane functions, and client service; inference is secondary and useful only while fresh. Deployment must therefore balance forecast freshness with network impact under contention.

For temporal forecasting, TimesFM and Toto~2.0 provide general baselines~\cite{timesfm,toto2}, TimeRAN develops lightweight foundation models for RAN analytics~\cite{timeran}, and APEX introduced network-native pre-training and compact variants evaluated on AP-class proxy hardware~\cite{apex}. Our work bridges these lines by fixing the forecasting task and defining \emph{network-aware deployability} as target-AP qualification of the model and execution path, followed by validation under packet-service and forecasting constraints. This approach separates model qualification from service admission and complements model-centric forecasting evaluations with constraint-aware systems validation. Overall, \emph{network-aware deployability} extends deployment evaluation beyond proxy performance to include target-device feasibility and safe coexistence with production networking.




\section{Conclusion}
Predictive inference on an enterprise AP is an \textit{asymmetric co-location} problem, not merely an \textit{edge-model optimization} problem. Our measurements show that the target and service feasibility are independent: a model and execution path qualified on the target device can still become service-infeasible under contention. Such workloads must therefore operate within a network-load-dependent envelope that protects networking and forecast freshness. Hence, the deployment objective is not just to deliver minimum inference latency but an optimal execution profile that preserves packet service while meeting forecast deadlines.